\documentclass[12pt,reqno,a4paper]{article}

\usepackage{amsmath, amssymb,graphicx,geometry,setspace,natbib}
\usepackage{hyperref}
\numberwithin{equation}{section}

\title{\bf Plankton: the paradox and the power law}

\author{Richard Law$^{1,\ast}$,
Jos{\'e} A. Cuesta$^{2,3}$,
Gustav W. Delius$^1$}

\date{}

\begin{document}

\maketitle

\noindent{}$^1$ Department of Mathematics and York Centre for Complex Systems Analysis,
              University of York, York, United Kingdom;

\noindent{}$^2$ Grupo Interdisciplinar de Sistemas Complejos (GISC) and UC3M-BS
              Institute of Financial Big Data (IFiBiD), Departamento de
              Matem\'aticas, Universidad Carlos III de Madrid, Madrid, Spain;

\noindent{}$^3$ Instituto de Biocomputaci\'on y F\'{\i}sica de Sistemas Complejos
              (BIFI), Universidad de Zaragoza, Zaragoza, Spain.

\noindent{}$\ast$ Corresponding author; e-mail: richard.law@york.ac.uk

\bigskip

\begin{abstract}

Two basic features of assemblages of unicellular plankton: (1) their high 
biodiversity and (2) the power-law structure of their abundance, can be 
explained by an allometric scaling of cell growth and mortality with respect 
to cell size. To show this, we describe a numerical study of a 
size-structured, multispecies, population-dynamic model; the model has a 
single resource, supporting an arbitrary number of phytoplankton and zooplankton 
species.  If the number of plankton species is large enough, the death rate of 
prey and cell growth rate of predators have approximate allometric scalings
with cell size.  Together, these scalings give rise to an equilibrium 
distribution of abundance near the power law, on which many species can
coexist.  Scalings of this kind cannot be achieved if the number of species 
is small. This suggests that the conjunction of species-richness and 
power-law structures in plankton communities is more than a coincidence.  
Although the exact allometric scalings used here should not be expected in 
practice, exclusion of species should be relatively slow if they lie close to 
the power law.  Thus the forces needed to achieve coexistence could be effective, 
even if they are relatively weak.


\end{abstract}

\section{Introduction}

This paper is motivated by two widely observed features of aquatic ecosystems.  
The first is a great diversity of plankton taxa, in the case of phytoplankton 
seemingly unconstrained by the small number of resources for which 
they compete (the paradox of the plankton: \citet{hutchinson:61}).  The 
second feature is a tendency for abundance of aquatic organisms to lie near a special 
power-law function of body mass.  This function corresponds to a biomass density 
of aquatic assemblages that changes rather little, as body mass (logarithmically 
scaled) is increased; it is found in the size range of the plankton \citep{gaedke:92, 
quinones:03, sanmartin:06}, and is thought to apply much more broadly from bacteria 
to whales \citep{sheldon:72}.  It appears difficult to account for both these features
simultaneously: they call both for a mechanism for species 
coexistence, and also for an emergent assemblage close to the power-law.
Although both observations have separately been the subject of much research, and 
the need to link them is well recognized \citep{armstrong:99}, a unified dynamical 
system that generates both features has not been described.

\begin{figure}
\begin{center}
\includegraphics[width=8cm]{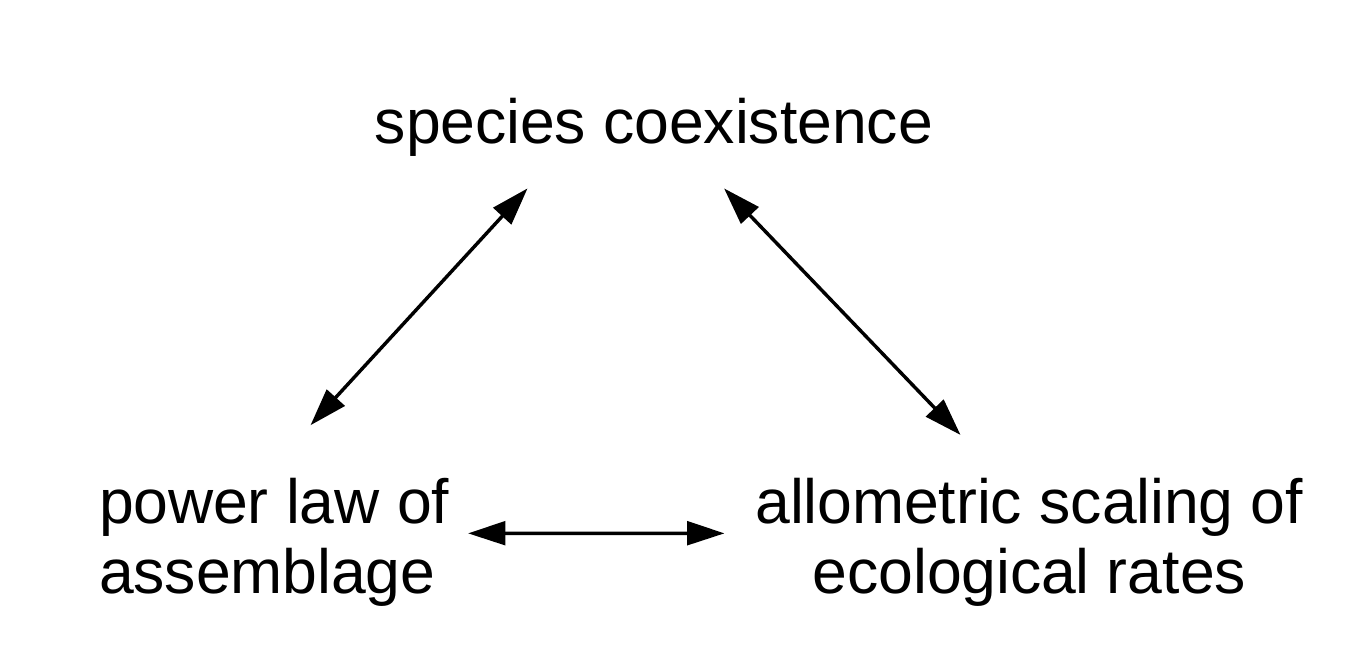}
\caption{Three basic observed properties of assemblages of unicellular plankton;
these motivate the multispecies, population-dynamic model in this paper.
}
\label{fig:triangle}
\end{center}
\end{figure}

In a phenomenological sense, coexistence of species is a prerequisite for the 
power law in assemblages of unicellular eukaryotic plankton.  These organisms 
only double in size before cell division, and consequently a power law 
spanning the size range of unicellular plankton depends on the coexistence 
of multiple species.  The purpose of the work here, and a mathematical paper 
that underpins it \citep{cuesta:16}, is to examine the {\it mechanistic} link 
between them.  For this we introduce a third ingredient:  allometric scalings 
between ecological rates and cell size.  These scalings are important because
without them a power-law solution would not be possible.  Some of the scalings, 
such those of metabolism and cell doubling time, are themselves well established 
empirically in unicellular plankton \citep{maranon:13, lopez:14}.  Other 
scalings are not, namely those of cell growth and death rates that stem 
from predation: these have to emerge directly from the predator-prey 
interactions.  The core of this paper is to show that predation can generate 
the allometric scalings needed, leading to results consistent with a triangle 
of linked observations shown in Fig. \ref{fig:triangle}.

Ecologists do not have a generally accepted mechanism for coexistence of a
large number of plankton species \citep{roy:07}.  Hutchinson's own preferred 
solution of environmental fluctuations favouring different species (the 
intermediate disturbance hypothesis) is insufficient in general \citep{fox:13}.  
It slows down the rate of loss of species, but this is not equivalent to placing 
bounds on abundance of species favouring them when they are rare and restricting 
them when common \citep{chesson:00, fox:13}.  One promising idea for bounding 
population increase in microbial ecology is ``killing the winner'', in which top-down 
control is strongest on the fastest-increasing prey \citep{thingstad:97, winter:10}.  
This was developed with phage and bacteria in mind, and parallels the idea of 
predator-mediated coexistence used elsewhere in ecology \citep{leibold:96, vage:14}.  
Here we extend the idea to predation in plankton assemblages.

Neither do ecologists have a clear-cut model mechanism to generate the observed 
power-law equilibrium in assemblages of unicellular plankton. Ideas are
better-developed in multicellular organisms from the study of so-called 
size-spectrum models \citep{silvert:78, silvert:80, benoit:04, 
andersen:06, capitan:10, datta:10, datta:11, hartvig:11, guiet:16}, where organisms 
grow, often over orders of magnitude, by feeding on and killing smaller 
organisms \citep{datta:10, plank:11}.  But the power-law structure extends 
down to unicellular plankton \citep{gaedke:92, quinones:03, sanmartin:06} where 
feedbacks are different because organisms do no more than double in body mass.
One suggestion is that, although plankton species may individually fluctuate 
in complicated ways, patterns become more regular after 
aggregation to the assemblage as a whole \citep{huisman:99, scheffer:03};
it remains to be seen whether a power-law pattern after aggregation 
is likely to emerge from this.

The work here builds on a mathematical theory proposed by \citet{cuesta:16} for 
scale-invariant dynamics of a continuum of species with a continuous trait 
(a characteristic body size) that spans an unlimited range from zero upwards, 
in addition to a continuous body size within species conventionally used in 
size-spectrum models. These so-called trait-based models are tractable 
enough to allow some formal mathematical analysis of multispecies assemblages 
\citep{andersen:06, hartvig:11}. However, they depend on assumptions: 
(1) that the trait is continuous (the number of species is infinite), and 
(2) that the trait is unbounded (i.e.\ the lower bound is zero, and the upper bound 
tends to infinity).  Clearly, real-world assemblages only support a 
finite number of species, and must have lower and upper bounds to cell size. 
This calls for a numerical analysis as given in this paper.

To keep a close link to the mathematical theory \citep{cuesta:16}, we retain its 
assumptions about allometric scalings of ecological rates with cell mass;  only
the scalings of cell death and growth are allowed to emerge from
the predator-prey interaction.  
However, it is not our intention to suggest that natural aquatic ecosystems satisfy 
the scalings exactly---there are many sources of variation that stand in 
the way of this. Our argument is that approximate scalings of processes with 
body mass in unicellular organisms bring them closer to ecological balance 
than has previously been thought. This slows down the rate of exclusion of 
species, with the consequence that other mechanisms for coexistence \citep{roy:07} 
can be more effective, even if their effects are not big.  Note also that the 
plankton communities here are built on a single limiting resource to show that
coexistence near the power law is possible even under the simplest and most 
exacting conditions;  we are not, however, suggesting that there is only one 
limiting resource in natural plankton assemblages.

\section{Theory}

We envisage an assemblage of unicellular plankton species, each species having a 
fixed characteristic cell size, defined as 1/2 of its maximum cell mass. For simplicity, 
mass is measured relative to that of a notional cell with equivalent spherical 
diameter (ESD) of 2 $\mu$m and mass $w_0$.  Throughout, we write $w$ as the 
mass expressed on this relative scale, and denote the characteristic
cell mass of species $i$ as $w_i$ on this relative scale.  Lower and 
upper bounds on cell size are set to span much of the range of unicellular 
eukaryotic plankton, with ESDs from approximately 2 to 200 $\mu$m.

The variables used are as follows.  $N$ is the concentration of nutrients
(mass volume$^{-1}$), and $P_i, Z_i$ are the total densities of a species of 
phytoplankton or zooplankton respectively, with species denoted by index $i$ 
(dimensions: volume$^{-1}$). However, to allow for allometric scalings in growth 
of cells, we disaggregate density down to cell mass within species.  The state 
variables are therefore $p_i(w), z_i(w)$ for the density of phytoplankton and 
zooplankton respectively at cell mass $w$ (dimensions also volume$^{-1}$,
because of the relative mass scale chosen above).  All these variables are 
time dependent, but for brevity the time argument is omitted.

The model depends on allometric scalings of cell growth and death 
with cell mass.  We therefore start with these details of cell life histories,
and use this as the input into the population dynamics of interacting species.  
In the population dynamics, phytoplankton species are coupled by competition
for a single limiting resource acting on the growth of cells.  Zooplankton 
and phytoplankton species are coupled by predation leading to death of 
phytoplankton cells and growth of zooplankton cells. Zooplankton species are 
coupled through predation by larger species on smaller ones.

\subsection{A scaling for growth of phytoplankton cells}

The von Bertalanffy growth model is a good place to begin, as it is based on
allometric scalings with body mass \citep{vonbertalanffy:57}.  It is more often 
thought of in the context of multicellular organisms, and is used here on 
the basis that, in unicellular organisms, growth still stems from the balance 
between (a) resource uptake, and (b) loss through metabolism  \citep{kempes:12}. 
For cell growth of a phytoplankton species $i$, we write 
$G^{(p)}_i(w)$: 
\begin{equation}
\frac{dw}{dt} = G^{(p)}_i(w) 
= \underbrace{A_iw^{\alpha}}_{(a)}
- \underbrace{B_iw^{\beta}}_{(b)} .
\label{eq:growth}
\end{equation}
Here $t$ is time, $\alpha$ and $\beta$ ($\alpha <\beta$) are parameters that scale 
respectively the gain of mass through resource uptake and the loss of
mass through metabolism with body mass $w$, and $A_i$ and $B_i$ are further 
species-dependent parameters for gain and loss respectively.   

It might seem from eq. \eqref{eq:growth} that there are two separate allometric
scalings with cell mass. However, it is known that over most of the size range of
eukaryotic phytoplankton species there is an approximate scaling of the intrinsic rate 
of increase \citep{maranon:13}.  This implies a single scaling of doubling time $T$ 
with respect to characteristic cell mass that has an exponent say $\xi$, i.e. $T \propto (w_i)^{\xi}$ 
(see definition of $w_i$ at start of this section).  The scaling of doubling time 
also implies a scaling on how fast cells from species with different $w_i$s grow: 
cells that take longer to double, must also have a lower mass-specific growth rate.  
Terms $A_i$ and $B_i$ in eq. \eqref{eq:growth} carry this dependence on $w_i$ and, 
with some algebra \citep{cuesta:16}, eq. \eqref{eq:growth} can be rewritten as:
\begin{equation}
G^{(p)}_i(w,N) = w^{1-\xi}\left[\tilde a(N)\left(\frac{w}{w_i}\right)^{\alpha+\xi-1}
-\tilde b\left(\frac{w}{w_i}\right)^{\beta+\xi-1}\right].
\label{eq:gp}
\end{equation}
where $A_i(N) = \tilde a(N)\cdot(w_i)^{-(\alpha+\xi-1)}$, 
$B_i = \tilde b\cdot(w_i)^{-(\beta+\xi-1)}$, $\tilde a(N)$ and $\tilde b$ are 
independent of 
$w_i$, and parameters $\alpha$ and $\beta$ are such that $2w_i$ is always less 
than the asymptotic value of cell size achievable from eq. \eqref{eq:gp}.  
The reason for writing the growth rate in this form is simply to make it clear
that all phytoplankton species have the same basic 
growth function.  This takes the form of a power function $w^{1-\xi}$ of cell mass, 
multiplied by a function of the ratio $w/w_i$.  In other words, the growth 
function scales with cell mass.  We have made use of a technical point that
allometric scaling can be done with respect to $w$ or $w_i$; see Appendix 
\ref{ref:w_wi_scaling}.

Note that, in eq. \eqref{eq:gp}, we have introduced a dependence of $\tilde a$ on 
the resource concentration $N$; $\tilde a$ is a strictly increasing function of 
$N$, since more resource leads to more gain in mass.  This function is 
important because it has the effect of coupling the dynamics of all phytoplankton 
species through depletion of the resource.

\subsection{A scaling for growth of zooplankton cells}
\label{ref:z_scalings}

We assume that zooplankton cells, like phytoplankton, grow through a balance between resource uptake
and metabolism, as in eq. \eqref{eq:growth}.  However, scaling of resource-uptake 
needs more attention because these cells grow by consuming other cells, 
and therefore depend on the abundance of these prey.  Thus, if this growth is to have an 
allometric scaling with cell size, it has to emerge from the predator-prey 
interactions.  Here we introduce three assumptions, justifiable on empirical 
grounds, that make an allometric scaling possible, although by no means inevitable.

The assumptions are as follows. (a) Zooplankton feeding depends on a volume sensed 
per unit time that scales with cell mass.  This is based on observations of 
encounters of protists with phytoplankton  \citep{delong:12}.  (b) Feeding 
occurs on cells 
around a fixed ratio of the mass of the consumer.  Size-based feeding is well
documented in plankton, although the predator-prey size ratios are smaller in
larger multicellular consumers \citep{wirtz:12}.  Within the size range of 
unicellular organisms, the change in ratio is relatively small, and a fixed 
ratio is a reasonable approximation \citep{wirtz:12}.  (c) Predation increases 
as prey species become abundant more than a simple law of mass action would allow;
this acts as a stabilizing force on the population dynamics \citep{chesson:00},
and is closely related to the notion of killing-the-winner 
\citep{thingstad:97, winter:10}.  These assumptions retain some elements of the 
complex food web that operates within plankton assemblages, avoiding the
reduction to a simple bulk phytoplankton--zooplankton trophic connection 
\citep{boyce:15}.

With these assumptions, the rate $S^{(p)}_{ij}(w,w')$ 
(respectively $S^{(z)}_{ij}(w,w')$) at which a zooplankton cell of species $i$ and 
size $w$ consumes a cell of species $j$ phytoplankton (respectively zooplankton) of size 
$w'$ ($w>w'$) is given by
\begin{equation} 
S^{(p)}_{ij}(w,w') = 
A w^{\nu} \times 
s(w/w') \times 
P_j^{\chi}
\label{eq:sp}
\end{equation}
\begin{equation} 
S^{(z)}_{ij}(w,w') = 
\underbrace{A w^{\nu}}_{(a)} \times 
\underbrace{s(w/w')}_{(b)} \times 
\underbrace{ Z_j^{\chi}}_{(c)} ,
\label{eq:sz}
\end{equation}
where (a), (b) and (c) formalize the three assumptions above.  In part (a), 
$A$ sets the overall level of encounters, and $\nu$ is the scaling exponent.  
The function (b) is a feeding kernel that distributes predation around a fixed
ratio of the predator cell mass, assumed for simplicity to be independent of
predator species $i$ and prey species $j$.  A positive value of $\chi$ in part (c) 
takes the mortality of prey species $j$ above the level of the law of mass action 
and would be expected to help stabilize the assemblage.  We used the total density 
of prey $j$ ($P_j$, $Z_j$) on the grounds that fine sorting of predation at the 
level of the size-disaggregated distribution would be unrealistic in organisms 
that only double in size.  This assumption could be changed with little effect 
on the results.

The growth rate $G^{(z)}_i(w)$ of a single zooplankton cell of species 
$i$ at mass $w$ now replaces the simple input term (a) in eq. 
\eqref{eq:growth} with the information on predation in eqs. \eqref{eq:sp}, 
\eqref{eq:sz}; the metabolic loss is unchanged.  This gives
\begin{eqnarray}
G^{(z)}_i(w) &=& \epsilon \int \sum_j w' S^{(p)}_{ij}(w,w') p_j(w') dw' \;\;\;\textrm{(gain from eating phytoplankton)}\nonumber\\
             &+&  \epsilon \int \sum_j w' S^{(z)}_{ij}(w,w') z_j(w') dw' \;\;\;\textrm{(gain from eating zooplankton)}\nonumber \\
             &-& w^{1-\xi} \;\tilde b\left(\frac{w}{w_i}\right)^{\beta+\xi-1} \;\;\;\textrm{(loss from metabolism)},
\label{eq:gz}
\end{eqnarray}
where $\epsilon$ is the efficiency with which prey mass is turned into predator mass.  
The gain terms have the form of a feeding rate at size $w$ on species $j$ at size $w'$, 
multiplied by the density of these prey and their cell mass $w'$, summed over 
prey species and integrated over prey cell masses, to get the total rate at 
which mass is consumed. 

A consistent allometric scaling of zooplankton growth requires that it should
match the scaling of metabolism, $w^{1-\xi}$, and hence the scaling of growth
in phytoplankton.  There is nothing in eq. \eqref{eq:gz} to ensure this happens. 
However, the scaling needed does emerge if the population densities lie on a 
power law, and we will show in Section \ref{ref:results} below that the predator-prey 
dynamics can generate this structure.  Such a power-law 
solution takes the form
\begin{equation}
 p_j(w)= w_j^{-\gamma} h(w/w_j) ,
 \label{eq:phat}
\end{equation}
where $\gamma$ is the power-law exponent.  Showing the statement is true is a 
matter of algebra, and we provide a proof in Appendix \ref{ref:z_growth_scaling};  
see also \citet{cuesta:16}.  The emergent scaling is important:  it means that 
there is a set of densities on a power-law of the form given in eq. \eqref{eq:phat} 
on which the growth rates of zooplankton species in eq. \eqref{eq:gz} have the same 
allometric scaling of cell mass, $w^{1-\xi}$, as phytoplankton species.  
This is necessary for the triangle of observations in Fig. \ref{fig:triangle} to 
emerge, although this does not preclude other outcomes that lack both the allometric 
scalings and also the power law (we will show an example in Fig. 
\ref{fig:loss_of_power-law}).  Note that the power-law exponent $\gamma$, expressed 
in terms of the other exponents $\nu$, $\xi$, $\chi$, is
\begin{equation}
 \gamma = 1+\frac{\nu+\xi}{1+\chi}  .
 \label{eq:power-law} 
\end{equation}
That is all we know so far as growth is concerned;  it is quite conceivable for 
a plankton assemblage neither to be on the power law, nor to relax onto it.

\subsection{A scaling for death rate}

We assume that most cell death comes from predation.  With one caveat, this
applies as much to zooplankton as to phytoplankton, because predators feed by cell
size rather than by cell type.  Cell death of prey is tied precisely to cell growth
of predators because each consumption event in eqs. \eqref{eq:sp}, \eqref{eq:sz} 
is matched by a corresponding prey death. Reversing the indices so that
$i$ is now a prey species of cell mass $w$, we write the rate $S^{(p)}_{ji}(w',w)$ 
(respectively $S^{(z)}_{ji}(w',w)$) as the rate at which a zooplankton cell of 
species $j$ and size $w'$ consumes a cell of species $i$ phytoplankton 
(respectively zooplankton) of size $w$ ($w'>w$).  Then the per-capita death rate 
from predation is part (a) of the following equations:
\begin{equation}
D_i^{(p)}(w) =  \int \sum_j S^{(p)}_{ji}(w',w) z_j(w') dw' + d_0 w^{-\xi} 
\label{eq:dp}
\end{equation}
\begin{equation}
D_i^{(z)}(w) =  
\underbrace{\int \sum_j S^{(z)}_{ji}(w',w) z_j(w') dw'}_{(a)} + 
\underbrace{d_0 w^{-\xi}}_{(b)} .
\label{eq:dz}
\end{equation}
In (a), the terms  $S_{ji}^{(.)}$ for consumption by $j$ are multiplied by the 
density of these predators.  The product is then summed over predator species, 
and integrated over predator cell masses, to get the total per-capita death rate 
from predation.

The caveat about predation is that, if this was the only cause of death, the
largest zooplankton would be free of all mortality; this is because we confine 
the study to unicellular organisms.  Mortality-free dynamics would be unrealistic, 
and we therefore introduce some background intrinsic mortality with the same 
scaling in all species, given as part (b) of eqs. \eqref{eq:dp}, \eqref{eq:dz}. 
Predation by multicellular organisms would remove the need 
for this, although some background mortality is to be expected in any event.
Thus the total per-capita death rate $D_i^{(p)}(w)$ of a phytoplankton cell 
(respectively $D_i^{(z)}(w)$ of a zooplankton cell) is given as the sum of 
(a) and (b) in eqs. \eqref{eq:dp}, \eqref{eq:dz}.

As in cell growth, there is an allometric scaling of predation with respect to $w$ 
when the population densities lie on the power law eq. \eqref{eq:phat}.  This can 
be shown by the method used for cell growth (Appendix \ref{ref:z_growth_scaling}),
and in this case the scaling is $w^{-\xi}$.  As noted for growth, the 
triangle of observations in Fig. \ref{fig:triangle} would not be possible 
without the scaling, although outcomes without this and without the
power law would be entirely feasible.

\subsection{Cell division rate}

In addition to cell growth and death, a cell-division process is needed for the
dynamics that follow.  This is essentially a matter of larger cells splitting
into two, but for realism, some variation in the size of daughter 
cells should be present immediately after cell division.  In general, there 
are two ways of generating this.  First, the rate of cell division in species 
$i$ can be a function $K_i^{(.)}(w)$ that becomes large as cell mass $w$ approaches 
$2w_i$. Secondly, there can be variation in size of the two daughter cells 
themselves.  For a cell that divides at size $w'$, 
we define a function $Q(w|w')$ that describes the probability density 
that the daughters are of size $w$ and $w'-w$, concentrated around $w = w'/2$.  
The two functions $K_i^{(.)}(w)$ and $Q(w|w')$ ensure that arbitrary features 
of initial size distributions decay over time in the population dynamics.

\subsection{A dynamical system for the NPZ assemblage}

\begin{figure}
\begin{center}
\includegraphics[width=12cm]{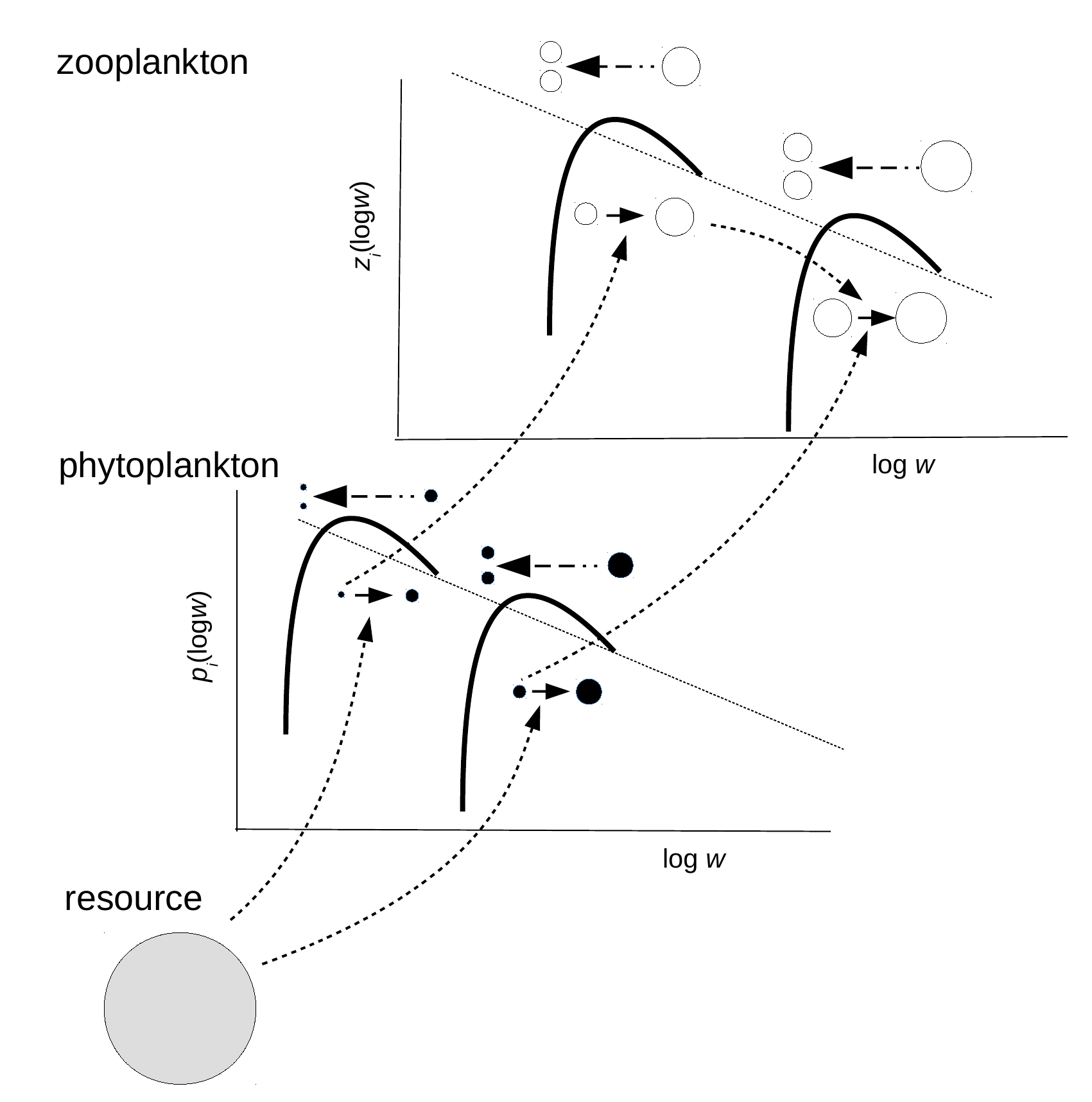}
\caption{Mass flows in an assemblage of unicellular plankton living on a single 
resource. Heavy lines are log-scaled size distributions of cells within species 
$\textrm{log} p_i(\textrm{log}w)$, $\textrm{log}z_i(\textrm{log}w)$ for phytoplankton 
and zooplankton species respectively.  
Continuous arrows denote growth of cells. Dotted arrows show the source of food
for cell growth; where growth comes from consuming smaller cells, this is accompanied
by death of these cells.  Dash-dot arrows denote cell division.
}
\label{fig:flows}
\end{center}
\end{figure}

So far, we have shown that an allometric scaling of cell growth and death with
$w$ is present, if population densities lie on a power law of the form in eq. 
\eqref{eq:phat}.  However, whether this special set of population densities could 
actually emerge from the predator-prey interactions is a separate matter, as yet 
unanswered.  This requires a model of multispecies population dynamics that 
can track the densities of cells over time. The model is built on the cell growth, 
death and division rates defined above, so that species are coupled through their 
predator-prey interactions and through their feeding on a common resource.
Fig. \ref{fig:flows} sketches the links involved.

The state variables for the model are the resource concentration $N$, the 
size distributions of an arbitrary number ($m$) of phytoplankton species $p_1(w), 
\dots, p_m(w)$, and the size distributions of an arbitrary number ($n$) of 
zooplankton species $z_1(w), \dots, z_n(w)$.

The exact form of the resource dynamics is not important, and we use a 
chemostat function
\begin{equation}
\frac{dN}{dt}=c\bigg(1 - \frac{N}{\hat N}\bigg) 
- \underbrace{d \sum_{i=1}^m\, \int_0^{\infty} A_i(N)\, w^{\alpha}\, p_i(w)\, dw}
_{\textrm{loss to phytoplankton}}.
\label{eq:resource}
\end{equation}
In the absence of phytoplankton, the resource settles to an equilibrium 
concentration $\hat N$.  The term for loss of resource to phytoplankton starts 
with the rate of resource uptake by cells of mass $w$ in species $i$ in
eq. \eqref{eq:growth}, integrates over all cell masses of species $i$
allowing for their densities, and then sums over phytoplankton species 
$1, \dots, m$.  The dimensionless parameter $d$ takes into account that 
growth is proportional to resource uptake, but not necessarily equal to 
it.  In this way, loss rate from the nutrient pool in eq. \eqref{eq:resource} 
matches the total uptake rate of resource, and couples together the 
dynamics of all phytoplankton species.

For the phytoplankton dynamics we use a form of the size-based 
McKendrick$-$von Foerster equation, which is essentially a cell growth-division 
equation.  This has a long history in cell biology \citep{fredrickson:67, 
bell:67, sinko:71, diekmann:83, Heijmans:84, henson:03, giometto:13}, and is an 
appropriate model for the dynamics of unicellular phytoplankton size distributions, 
shown here for species $i$:
\begin{eqnarray}
\frac{\partial}{\partial t}p_i(w) =
&-&\frac{\partial}{\partial w}\big[G^{(p)}_i(w,N)p_i(w)\big] \nonumber \\
&+&2\int_0^\infty Q(w|w')K_i^{(p)}(w')p_i(w') dw' \nonumber \\
&-&K_i^{(p)}(w)p_i(w) \nonumber \\
&-&D_i^{(p)}(w)p_i(w) ,
\label{eq:MvFP}
\end{eqnarray}
with $G_i^{(p)}$ from eq. \eqref{eq:gp}, and $D_i^{(p)}$ from eq. 
\eqref{eq:dp}.  The first term on the right-hand side is the rate at which 
cell growth leads to change in density at size $w$.  The second term is 
the rate at which new cells are generated at size $w$ from cell
division; the probability density $Q$ concentrates the new cells around 
$w'/2$. The third term is the rate at which cells disappear at size $w$ through 
cell division, $K_i^{(p)}$ making this term large near $2w_i$. The fourth 
term is the rate of cell loss from mortality.  Similarly the zooplankton 
dynamics for species $i$ are given by
\begin{eqnarray}
\frac{\partial}{\partial t}z_i(w) =
&-&\frac{\partial}{\partial w}\big[G^{(z)}_i(w,N)z_i(w)\big] \nonumber \\
&+&2\int_0^\infty Q(w|w')K_i^{(z)}(w')z_i(w') dw' \nonumber \\
&-&K_i^{(z)}(w)z_i(w) \nonumber \\
&-&D_i^{(z)}(w)z_i(w).
\label{eq:MvFZ}
\end{eqnarray}
with $G_i^{(z)}$ as in eq. \eqref{eq:gz}, and $D_i^{(z)}$ as in eq. 
\eqref{eq:dz}.

\section{Results on the NPZ equilibrium}
\label{ref:results}

\subsection{Continuum model}

The continuum model allows some formal mathematical results to be given about 
the steady-state behaviour \citep{cuesta:16}.  This model replaces the finite
number of species in a bounded range of body size in eqs. \eqref{eq:MvFP}, 
\eqref{eq:MvFZ}, with a continuum of species that have characteristic cell masses 
$w_*$ spanning the range from zero to infinity. Remarkably, it can be proved that 
this model has a power-law equilibrium.  This is an equilibrium on which
an infinite number of species coexist, notwithstanding the competition for
resource among the phytoplankton and the predator-prey interactions 
between zooplankton and phytoplankton.  The equilibrium has the form 
given in eq. \eqref{eq:phat}, and the abundance of the characteristic 
masses $w_*$ scales with $w_*$ with the exponent given in eq. 
\eqref{eq:power-law}. In this case, the equations and the steady-state solution 
are scale invariant.  

The unique power-law equilibrium emerges from the dynamics as a direct consequence 
of the predator-prey interactions.  The power-law structure of the zooplankton holds 
death rates from predation with a scaling exponent $-\xi$, consistent with the 
power law of prey abundance.  At the same time, the power-law structure 
of the prey holds the growth rates of the zooplankton with a scaling exponent 
$1-\xi$, consistent with the power law of predator abundance.  Whether this 
equilibrium could be stable without additional stabilising mechanisms is not known.

\subsection {Discretized model}

For two reasons, it helps to go from the continuum model to a numerical 
analysis of the discretized, bounded system. First, the number of species 
is always finite in reality, and secondly there must always be lower and upper 
bounds on cell size in the plankton.  In these circumstances, the perfect 
power-law equilibrium in the idealised continuum model \citep{cuesta:16} can 
never be achieved exactly, and a numerical analysis can show whether it is a 
useful guide to more realistic systems.  In addition, a numerical stability 
analysis can show whether the near-power-law equilibrium is an attractor, 
in other words, whether plankton assemblages should actually be expected to 
move towards it;  this has not been possible in the continuum model.
The numerical analysis requires some functions and parameter values to 
be specified;  these are given in Table \ref{table:functions} and Table 
\ref{table:parameters} in Appendix \ref{ref:parameter_values}.

\begin{figure}
\begin{center}
\includegraphics[width=14cm]{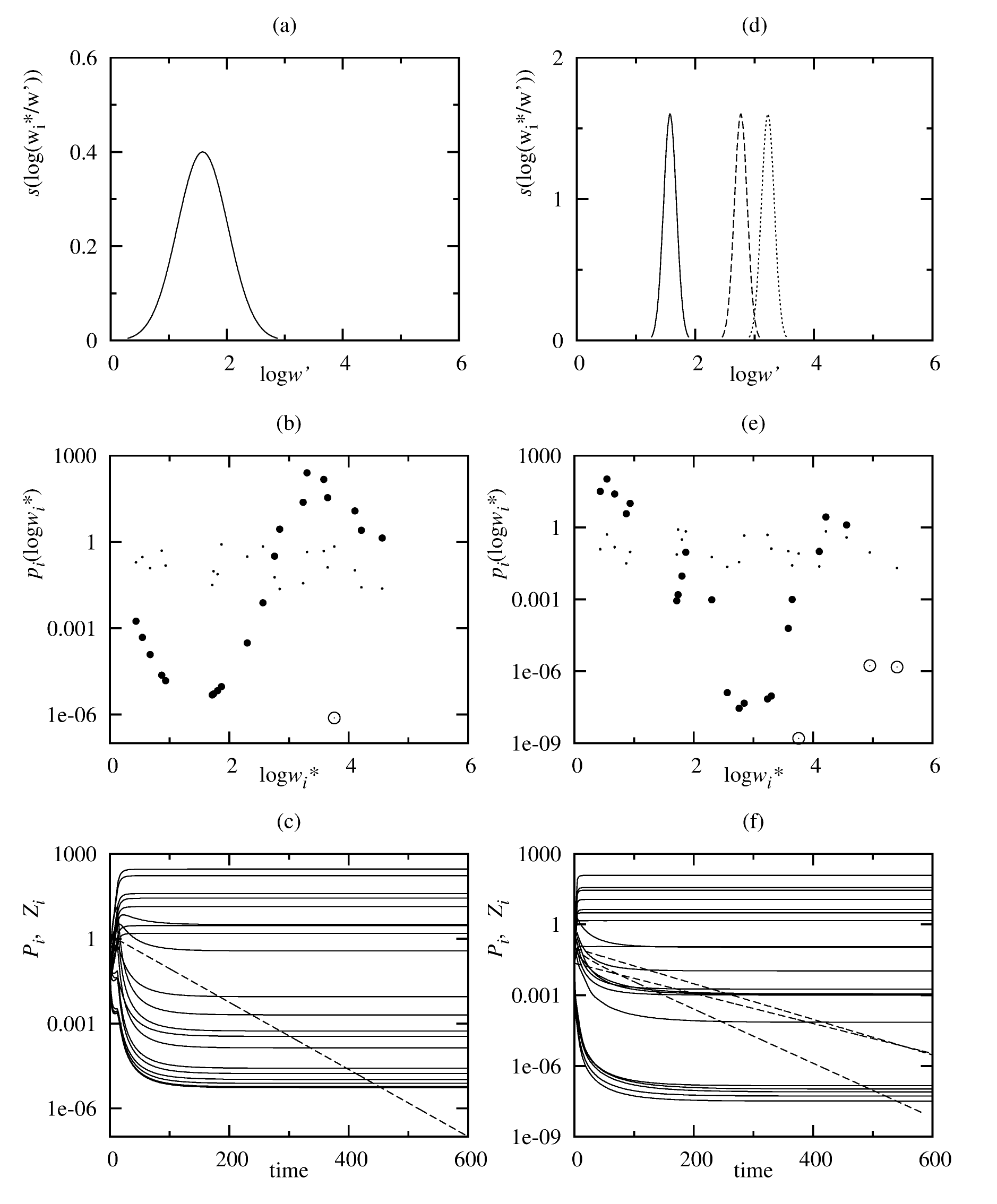}
\caption{Absence of a power-law relationship between abundance and cell mass
in an assemblage of 20 phytoplankton and a small number of zooplankton species.  
(a), (b), (c) One zooplankton species with diet breadth $\sigma = 1.0$. 
(d), (e), (f), Three zooplankton species with diet breadths $\sigma = 0.25$.
(a), (d) Feeding kernels for zooplankton cells at characteristic cell sizes
$w_i$ of each species.  
(b), (e) Size spectra for $w_i$ at time 600 of phytoplankton (filled circles), 
and zooplankton (open circles) species;  the initial densities are shown as dots.  
(c), (f) Time series of the total densities of phytoplankton (continuous) and 
zooplankton (dashed) species.  The time series were obtained from
numerical integration of eqs. \eqref{eq:resource}, \eqref{eq:MvFP},
\eqref{eq:MvFZ}, with $a(N)$, $Q(w|w')$, $K_i^{(.)}(w)$ and $s(w/w')$ as in Table 
\ref{table:functions}, and parameter values as in Table \ref{table:parameters}
in Appendix \ref{ref:parameter_values}.
}
\label{fig:loss_of_power-law}
\end{center}
\end{figure}

Note first that the NPZ dynamical system eqs. \eqref{eq:resource}, 
\eqref{eq:MvFP}, \eqref{eq:MvFZ} can reach an equilibrium far away from any
power law (Fig. \ref{fig:loss_of_power-law}).  This is clear if, for instance, there is a just a single 
zooplankton species, restricting the size range over which predation 
takes place; Fig. \ref{fig:loss_of_power-law}a shows the feeding kernel at 
the characteristic cell mass of this species.  The densities of vulnerable 
phytoplankton species are driven down to low densities by this localised 
feeding of the predator, creating a corresponding `hole' in the phytoplankton 
assemblage (Fig. \ref{fig:loss_of_power-law}b). This leaves the zooplankton 
species without enough food, and it tends to zero density (Fig. 
\ref{fig:loss_of_power-law}c).  As the predator goes to extinction, the phytoplankton 
assemblage comes to rest at a point where the most vulnerable species 
have densities close to zero.  Put another way, with only one zooplankton species,
the predator-prey interaction does not allow the predation-death to settle 
near an allometric scaling near $-\xi$.  More specialised feeding exacerbates the 
problem, as can be seen in Fig. \ref{fig:loss_of_power-law}d,e,f, where there 
were initially three zooplankton species with narrow feeding kernels.
Because the abundances are nowhere near a power law, this behaviour is not 
consistent with the full triangle of observations in Fig. \ref{fig:triangle}.  
This is despite the fact that the phytoplankton species do coexist without 
the predator, and do have an allometric scaling of growth and death rates 
with respect to cell mass.  The allometric scaling of death comes from the background
mortality that remains in eq. \eqref{eq:dp} in the absence of predation.

The key numerical result of this paper is that a larger number of 
zooplankton species brings death (and cell growth) from predation 
approximately to the allometric scaling with cell mass needed for the power law. 
In effect, more zooplankton species distribute predation mortality better across 
the range of cell sizes.  The outcome is then consistent with the 
full triangle of observations (Fig. \ref{fig:triangle}), namely: coexistence 
of species, allometric scalings of ecological rates, and the power-law 
structure of the assemblage.  Moreover, the numerical results show that the 
equilibrium can be locally asymptotically stable, given a sufficiently 
positive value of $\chi$.  In other words, there is a neighbourhood 
of the near-power-law equilibrium within which all other initial size 
distributions return to the equilibrium.  The equilibrium is at least a 
local attractor.   

\begin{figure}
\begin{center}
\includegraphics[width=15cm]{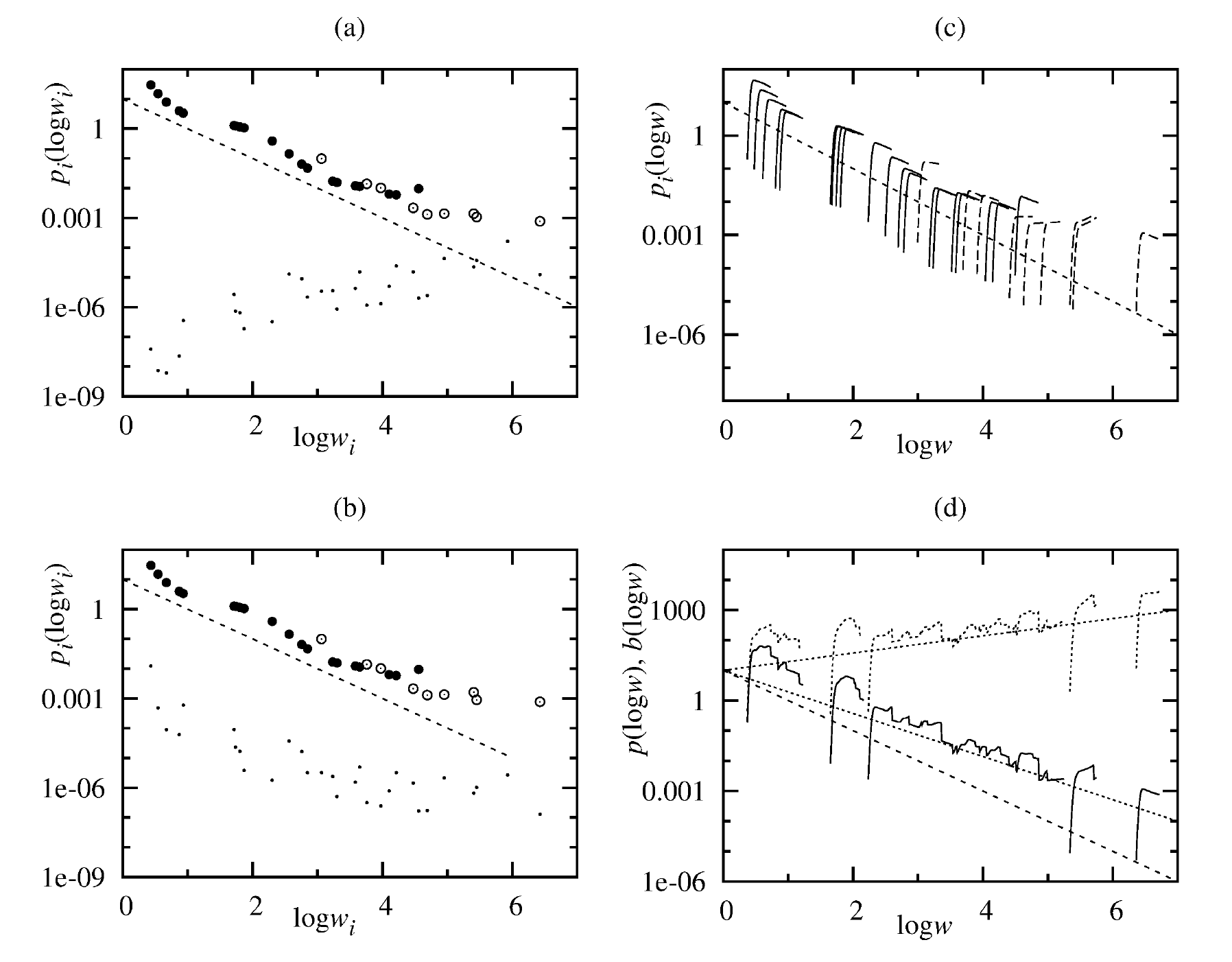}
\caption{Convergence to a near-power-law equilibrium of a plankton community of 
20 phytoplankton species in presence of predation by 9 zooplankton species.
(a), (b) Two contrasting initial conditions lead to the same equilibrium densities
of $w_i$ (characteristic cell masses) of phytoplankton (filled circles), and 
zooplankton (open circles) species;  the initial densities are shown as dots. 
(c) Complete size distributions within phytoplankton (continuous) 
and zooplankton (dashed) species at equilibrium.
(d) Community spectra aggregated over species at equilibrium: size spectrum 
$p(\textrm{log}w)=\sum_i p_i(\textrm{log}w)$ (continuous line), biomass spectrum 
$b(\textrm{log}w)= \sum_i wp_i(\textrm{log}w)$ 
(dotted line).  
The dashed reference lines have a slope -1, and the dotted lines have slopes 
2/7 and -5/7, corresponding to eq. \eqref{eq:power-law} after transforming 
from $w$ to $\textrm{log}w$. 
The set of phytoplankton species and the first three zooplankton species from 
Fig. \ref{fig:loss_of_power-law} were used, together with seven further
zooplankton species.  
Numerical integrations were carried out as in Fig. \ref{fig:loss_of_power-law} 
with parameter values as in Table \ref{table:parameters}.  Integration brought 
the densities close to equiibrium, after which the equilibrium was obtained by 
Newton's method.  The Jacobian at equilibrium had a leading eigenvalue with
a real part = -0.0018, implying local asymptotic stability of the equilibrium.
}
\label{fig:power-laws_PZ}
\end{center}
\end{figure}

We illustrate this result in Fig. \ref{fig:power-laws_PZ}, in which we took an 
assemblage of 20 phytoplankton and 10 zooplankton species with random 
characteristic cell masses. The second largest zooplankton species tended 
to zero density, and the 29-species equilibrium (with the second-largest species 
at zero density) was locally asymptotically stable.  Importantly, the
equilibrium was quite close to a power law (i.e. close to linear in the double
logarithmic plot);  this is shown at the characteristic cell masses 
(Fig. \ref{fig:power-laws_PZ}a,b), and for the full size distributions of 
the species (Fig. \ref{fig:power-laws_PZ}c).  It is notable 
that the two illustrative initial conditions in Fig. \ref{fig:power-laws_PZ}a,b 
are far from the equilibrium point, suggesting that the equilibrium has a
basin of attraction substantially greater than the range of initial conditions
corresponding to a local stability analysis.    
This ordered structure is generated by the predator-prey interaction.  
A slope of $2/7$ for the biomass spectrum is predicted from 
eq. \eqref{eq:power-law}, and the spectrum that emerged from the numerics is 
close to this, especially when away from the lower and upper boundaries of 
cell size (Fig. \ref{fig:power-laws_PZ}d). The lack of predation on 
species near the upper boundary would be expected to leave a relatively
large footprint near the right-hand boundary, and this is indeed evident 
in the size distributions and overall densities of these species (Fig. 
\ref{fig:power-laws_PZ}c).

\section{Discussion}

This study shows, for the first time, how predator-prey dynamics can
drive unicellular plankton assemblages towards a power-law relationship
between cell mass and abundance.  With enough species, the near-power-law 
equilibrium can be an attractor, and the predator-prey interaction can bring the 
whole assemblage to this state.  In this way we obtain a mechanism for
coexistence of multiple plankton species near a power-law equilibrium, consistent
with the triangle of observations in Fig. \ref{fig:triangle}.  The study gives 
some support to the conclusions of a mathematical analysis of the continuum model,
in which an unlimited number of unicellular plankton species can coexist 
at a power-law equilibrium if cell growth predation and death 
all scale appropriately with cell mass \citep{cuesta:16}.   

The size-based modelling of predation adopted here improves the resolution of 
trophic structure in plankton assemblages, as called for by \citet{boyce:15}.  It 
would readily extend to cover mixotrophic feeding \citep{flynn:13}.  Also, a 
straightforward extension would be to move the upper bound on body mass to 
encompass metazoan plankton and fish, and we would expect this to remove some of the 
roughness at the right-hand side of the plankton spectrum in our computations.  
Taking this size-based approach through to larger organisms would help to 
provide a common formal framework for the flow of biomass more broadly 
through aquatic ecosystems, so-called end-to-end models \citep{fulton:10}, 
eliminating the need for closure assumptions at the upper end of the plankton 
\citep[e.g.][]{edwards:01} and at the lower end of fish assemblages 
\citep[e.g.][]{datta:10, hartvig:11}.  The extension to 
more than one limiting resource would not change the qualitative picture 
we are presenting here.


The caveat about the continuum (mathematical) model \citep{cuesta:16} is that, 
although it shows the power-law equilibrium exists, as it stands this equilibrium 
is not known to be an attractor.  However, one further assumption can make 
the equilibrium stable to small displacements;  this is that the dependence of 
predation on the density of prey is stronger than that given by the standard 
rule of mass-action.  The effect is to make over-abundant species experience 
disproportionately large mortality.  This is closely related to the idea of 
killing-the-winner in which prey (hosts) are controlled according to how 
fast they are increasing \citep{thingstad:97, winter:10}.
With this assumption in place, the exponents in the model are related by 
$\gamma = 1 + (\nu+\xi)/(1+\chi)$, and $-\gamma$ emerges as 
the exponent of the power-law relationship between abundance and 
characteristic cell mass.

Such predation mortality could come about in 
various ways, for instance through the ideal-free distribution 
\citep{kacelnic:92}, redistribution of predation according to prey species 
abundance \citep{armstrong:99}, patchy spatial distributions of prey 
coupled to predator aggregation \citep{nachman:06}, and vertical profiles 
of phytoplankton and zooplankton in the water column \citep{morosov:10}.
A Holling type III functional response \citep{holling:59} has been suggested 
as a surrogate for such spatial effects \citep{nachman:06, morosov:10}, and 
was needed to generate plankton blooms in excitable-media models 
\citep{truscott:94}.  The predation mortality has the effect that predators feed 
disproportionately on prey species with an excess of biomass, thereby 
reducing the size of peaks and troughs in the community biomass spectrum. 
This is not to discount the possible existence of other mechanisms that 
could stabilize the equilibrium.


An important feature of the near-power-law equilibrium abundance is that it 
needs enough species to be present to generate allometric scalings
for prey death and predator growth. The continuum (mathematical) model 
shows that the power-law equilibrium becomes exact as the number of species tends to 
infinity \citep{cuesta:16}, and this suggests 
that the allometric scalings become easier to achieve as the number of prey 
and predator species increases.  Broader diets, i.e. feeding over a greater 
range of prey sizes (coincidentally increasing connectance), would also help 
in generating the scalings.  The coexistence of many species at equilibrium near 
the power law is not a consequence of more feeding niches with increasing 
species richness; there is, for instance, no lack of niche space for the 
few predator species in Fig. \ref{fig:loss_of_power-law} at time 0.  
Neither is there a limiting similarity caused by niche overlap: species 
can be arbitrarily close together at the power-law equilibrium of the 
underlying continuum theory \citep{cuesta:16}. Coexistence of such similar
species is consistent with recent molecular evidence of coexisting cryptic species 
\citep[e.g.][]{amato:07, mcmanus:09}. 

These features of the power-law equilibrium are unanticipated in the context 
of standard ecological theory.  This theory does not lead to an expectation that 
an unlimited number of species could coexist, unconstrained by food-web 
connectance, and without the need for separate niches.  In particular, a great 
deal of research on random matrices from \citet{may:72} onwards leads to the 
view that ecological communities are less likely to be stable as species 
richness and connectance increase.  We think it likely that there is 
structure in assemblages of aquatic plankton that random matrices have yet 
to take into account.  Random matrices would need to generate equilibria that
come close to a power-law relationship with cell size, and this probably 
requires appropriate allometric scalings of ecological rates.


Note that the study of species-poor plankton communities gives no hint that
coexistence could be readily achieved in species-rich communities. 
This invites an obvious and important question as to how a species-rich 
community close to the power-law abundance could be assembled from its 
component parts in the first place.  That such assemblages exist is not in doubt 
\citep{gaedke:92, quinones:03, sanmartin:06}.  However, it is notable that 
the biomass spectra of some lake communities have marked peaks and troughs, 
unlike oligotrophic ocean systems \citep{sprules:91, quinones:03, yurista:14},
suggestive of an intermediate state in assembling plankton communities.  
Assembly is an interesting matter that needs further research.


Clearly, many real-world processes prevent the existence of perfect power-law 
size spectra and exact allometric scalings of ecological rates.  External 
processes include seasonal fluctuations that leave a strong footprint in 
community size spectra \citep{heath:95}.  Exploitation of fish stocks 
causes major changes in size spectra further up the food web \citep{blanchard:05}, 
which may be felt lower down through trophic coupling in the ecosystem.  
Internal processes include species-specific features of plankton life histories 
that a simple generic model cannot incorporate.  For instance, the characteristic cell 
mass of diatoms decreases as they go through cycles of asexual reproduction; 
some taxonomic specificity in feeding has been observed \citep{jezbera:06} and there 
is evidence that unicellular taxonomic groups have different preferred predator-prey 
size ratios \citep{hansen:94}. The allometric scaling relationships may also change;
for instance \citet{maranon:13} found a positive relationship between intrinsic rate
of increase and cell size below cell diameters $\simeq 5$ $\mu$m, and there is some
change in the scaling of optimal prey size from small to large planktonic predators
\citep{wirtz:12}.

Bearing in mind these caveats, the strongest statement this paper can support 
is that unicellular plankton species are more balanced in their interactions when close to the power 
law, than when further away from it.  This means that the forces driving exclusion 
of one species by another are likely to be weaker than has previously been 
thought. Correspondingly, the forces needed to counteract exclusion and to maintain 
species-rich plankton assemblages do not have to be so strong to work effectively.
A corollary is that this balanced state of coexistence can be disrupted by processes,
human or otherwise, that drive aquatic ecosystems further from the power law.

\section {Acknowledgements}

RL and GWD were supported by EU Grant 634495 — MINOUW — H2020-SFS-2014-2015. 
JAC was supported by the Spanish mobility grant PRX12/00124 and project 
FIS2015-64349-P (MINECO/FEDER, UE).  We thank A. D. Dean, J. W. Fox, J. Kolding, 
E. Mara{\~n\'o}n, E. J. A. Minter, M. J. Plank and S. Va\r{g}e for helpful 
discussions on this work.

\appendix
\section*{Appendices}

\section{Allometric scaling and power-law abundances}
\label{ref:w_wi_scaling}

Allometric scaling of physiological rates with cell size plays an essential role 
in this paper.  There is however a potential for confusion as to whether the 
allometric scaling depends on the cell size $w$ of an individual or on 
the characteristic cell size $w_i$ of a species. We clarify this by acknowledging
that rates can depend on both $w$ and $w_i$, and that allometric scaling refers 
to the behaviour of the rate under a rescaling of both $w$ and $w_i$. 
If $R_i(w)=R(w,w_i)$ denotes some physiological
rate for species $i$, then allometric scaling with an exponent $\rho$ means that
the rate $R(w,w_i)$ satisfies
\begin{equation}
R(\lambda w,\lambda w_i) = \lambda^\rho R(w,w_i)
\end{equation}
for any positive scale factor $\lambda$.  In other words, $R(w,w_i)$ is a 
homogeneous function of both arguments. 
This in turn means that the rate can be written in two alternative scaling forms
\begin{equation}
R_i(w)=w_i^\rho \,r(w/w_i) =w^\rho\, \tilde{r}(w/w_i)
\end{equation}
where $\tilde{r}(x)=x^{-\rho} r(x)$. Thus it 
does not matter whether one thinks of allometric scaling in terms of cell size $w$ or in terms of characteristic
size $w_i$. The exponent will be the same in either case and only the scaling function
of $w/w_i$ will differ.

A similar issue also arises when abundance power laws are discussed: are these 
power laws in cell size $w$, or in the characteristic cell size of a species 
$w_i$?  Again it helps to view the abundance as a function of both $w$ and $w_i$, 
as we do in this paper. Saying that the abundance has a power-law form then means 
that it is a homogeneous function of both arguments and thus can be written 
equally well in terms of a power of $w$ or a power of $w_i$. For example in the 
case of the steady state phytoplankton abundance it is equally valid to write 
$p_i(w) = w_i^{-\gamma-1}f_p(w/w_i)$ as it is to write
$p_i(w) = w^{-\gamma-1}\tilde{f}_p(w/w_i)$.

\section{Allometric scaling of zooplankton growth}
\label{ref:z_growth_scaling}

Here we show that there is an allometric scaling of zooplankton growth 
$G_i^{(z)}(w)$ of the form $w^{1-\xi}$ in eq. \eqref{eq:gz}, if a multispecies
plankton assemblage is at a power-law equilibrium.  The argument begins by assuming 
a power-law equilibrium with densities $p_j(w_j)$ (respectively $z_j(w_j)$) at 
$w_j$ that scale with characteristic cell mass as $w_j^{-\gamma}$, where 
$\gamma$ is an unknown exponent.  Also assume that the equilibrium densities 
at sizes other than $w_j$ can be written as $p_j(w)= w_j^{-\gamma} h(w/w_j).$

Make a change in variable $v=w'/w_j$ in the predation terms of eq. 
\eqref{eq:gz}, giving $w'=w_jv$ and $dw'=w_jdv$; this allows $w_j^2$ 
to be factored out of the predation integrals.  
Substituting eq. \eqref{eq:phat} into \eqref{eq:gz}, allows a further 
$w_j^{-\gamma}$ to be factored out.  The same factorisations apply to $z_j(w')$. 
Using eq. \eqref{eq:phat} gives 
$P_j^\chi = w_j^{(1-\gamma)\chi}(\int h(v) dv)^\chi$; 
when substituted into eq. \eqref{eq:gz}, this allows a further 
$w_j^{(1-\gamma)\chi}$ to be factored out.
The same factorisation applies to $Z_j^\chi$. 
Lastly, $w^\nu$ is factored out of the predation integrals.  
Noting that $w_i^\zeta$ can always be written in the form 
$w^\zeta (w/w_i)^{-\zeta}$, means that the exponents can be collected 
together as powers of $w$, namely $2-\gamma+(1-\gamma)\chi+\nu$. 
This function of exponents can now be equated with the allometric scaling of 
metabolism $1-\xi$ so that the $1-\xi$ scaling applies throughout $G_i^{(z)}$.

\section{Parameter values}
\label{ref:parameter_values}

The characteristic cell mass $w_i$ of species $i$ is defined as half its maximum 
cell mass. The discrete set of $w_i$s replace the continuum of species with 
characteristic cell mass $w_*$ in \citet{cuesta:16}. In phytoplankton, values 
for $w_i$ were chosen in the range $2 \rightarrow 30000$ pg, corresponding to 
upper limits on cell ESDs of $\sim 2 \rightarrow 40 \mu$m.  Those
of zooplankton were chosen in the range $200 \rightarrow 3 \times 10^6$ pg, 
corresponding to upper ESD limits of $\sim 10 \rightarrow 200 \mu$m.  Staggering the cell size 
ranges in this way ensured that all phytoplankton were vulnerable to predation, and
all zooplankton had food to eat.

From a compilation of earlier studies, \citet[][Fig. 2]{tang:95} estimated
algal cell division rate to scale with an exponent of about $-0.15$ with respect
to cell size.  Recently, it has been noted that cell division rates reach
a maximum at a cell diameter $\simeq 5$ $\mu$m \citep{chen:10, maranon:13}; the
change is particularly clear in the transition from bacteria to eukaryotes
\citep{kempes:12}.  \citet{maranon:13} gave an exponent for the rate of 
increase of cells on the right-hand side of the peak of approximately $ -0.15$, 
corresponding to a cell doubling time scaling as $\xi = 0.15$, the value used 
here. 
 
Metabolism was assumed to scale isometrically with cell mass $\beta = 1$, 
in keeping with observations on protists in \citet{delong:10} and phytoplankton
\citep{lopez:14}.  The rate of gain in cell mass was assumed to scale with an exponent 
$\alpha = 0.85$.  With the values chosen, the structure of eq. \eqref{eq:growth} is 
simple, as $\alpha+\xi-1 = 0$.  Note that $\alpha$ must be less than $\beta$ to 
ensure that metabolic loss eventually becomes greater than the gain in mass, as 
cells grow.

\begin{table}
\caption{Explicit functions} 
\begin{center}
\begin{tabular}{l l l p{6.5cm}}

Name &Eq. &Function &Comments\\

\vspace{0.1 cm}\\
$ \tilde a(N) $    & \eqref{eq:gp}     
& $\frac{\tilde a_\infty N}{r+N}$   
& functional response for resource uptake\\

\vspace{0.1 cm}\\
$ K_i^{(.)}(w)$   & \eqref{eq:MvFP}, \eqref{eq:MvFZ}  
& $G_i^{(.)}(w)\delta(w-2w_i)$ 
& cell division rate set to zero, except at $2w_i$ ($\delta$ is the Dirac $\delta$-function)\\

\vspace{0.1 cm}\\
$ Q(w|2w_i) $ & \eqref{eq:MvFP}, \eqref{eq:MvFZ}  
& $\frac{1}{\sqrt{2\pi}\sigma_bw_i} \exp\big(-\frac{(w/w_i-1)^2}{2\sigma_b^2}\big)$ 
& distribution of cell masses following division at $2w_i$ \\

\vspace{0.1 cm}\\
$ s(w/w')$  & \eqref{eq:sp}, \eqref{eq:sz}  
&$\frac{1}{\sigma \sqrt{2\pi}} \exp\big(-\frac{(\ln (w/w')-\Delta)^2}{2\sigma^2}\big)$ 
& feeding kernel\\

\end{tabular}
\end{center} 
\label{table:functions}
\end{table}

\begin{table}
\caption{Model parameters and values used in Figs. \ref{fig:loss_of_power-law} and
\ref{fig:power-laws_PZ}}
\begin{center}
\begin{tabular}{l l l l l l p{6.0cm}}

\vspace{0.01 cm}
Symbol & Value & Dimensions & Comments\\

\vspace{0.01 cm}\\
$ \xi $      & 0.15 &--       & exponent scaling doubling time with cell mass\\

\vspace{0.01 cm}\\
&&&\textit{Metabolic loss}\\
$ \beta $    & 1    &--       & exponent scaling metabolism with cell mass\\
$ \tilde b $ & 0.5  &T$^{-1}$ & metabolism term\\

\vspace{0.01 cm}\\
&&&\textit{Resource uptake}\\
$ \alpha$           & 0.85 &--       & exponent scaling resource uptake with cell mass\\
$ \tilde a_\infty$  & 2    &T$^{-1}$ & resource uptake parameter\\

\vspace{0.01 cm}\\
&&&\textit{Intrinsic death rate}\\
$ d_0 $      & 0.1  &T$^{-1}$ & intrinsic death rate parameter\\ 

\vspace{0.01 cm}\\
&&&\textit{Cell division}\\
$\sigma_b$   & 0.05 & --      & size range of daughter cells\\

\vspace{0.01 cm}\\
&&&\textit{Resource dynamics}\\
$ c $        & 100 & M V$^{-1}$ T$^{-1}$ & resource growth rate\\
$ \hat N $   & 100 & V$^{-1}$            & resource equilibrium without plankton\\
$ d $        & 1   & --                  & proportionality constant for resource uptake to cell growth\\
$ r $        & 1   & V$^{-1}$            & type II functional response term\\

\vspace{0.01 cm}\\
&&&\textit{Predation}\\
$ A $        & 0.02   & V$^{-1}$ T$^{-1}$ & encounter parameter\\
$ \nu $      & 0.85   & -- & exponent volume sensed\\
$ \Delta $   & 5      & -- & log preferred predator prey mass ratio\\
$ \sigma $   & 1, 0.25& -- & diet breadth\\
$ \epsilon$  & 0.6    & -- & food conversion efficiency\\
$ \chi $     & 0.4    & -- & exponent for density dependence\\

\vspace{0.01 cm}\\
&&&\textit{Numerics}\\
$ \delta x $ & 0.025  & -- & size step (logarithmic binning)\\
$ \delta t $ & 0.0005 & T  & time step\\
$ t_{max} $  & 600    & T  & time period for integration\\

\end{tabular}
\end{center} 
\label{table:parameters}
\end{table}

We set $\tilde a_{\infty}$ in the function $\tilde a(N)$ (Table \ref{table:functions}) 
to be substantially greater than $\tilde b$ (2, 0.5 respectively) so that 
phytoplankton cells would not shrink when the resource level was below 
its equilibrium level. It was still possible for there to be insufficient resource
for phytoplankton cells to grow, in numerical integrations starting at high cell 
densities. The same could apply to zooplankton cells if the food available was
too low.  To deal with such cases, negative growth rates were replaced by zero, 
causing cells to stay at the same size, experiencing the level of mortality 
corresponding to that size.  However, at the equilibrium point itself, the 
growth rate is positive in every species with a positive density. This is 
intuitive.  If it was not so, cells would not get as far as cell division 
and the species concerned would be decreasing in density, not at equilibrium.
 
The intrinsic death parameter $d_0$ was set at 0.1, so that most mortality 
would be caused by predation.  Some residual mortality is needed for the 
largest zooplankton species {\it in lieu} of larger predators; for 
compararability, all species were given the same background value.

The parameter $\sigma_b$ spreads out the size of daughter cells. We 
investigated the effects of this and chose $\sigma_b = 0.05$ as ensuring 
enough variation for effects of initial size structure within species to 
attenuate quickly.
 
Parameter $A$ describes the overall level of encounters of zooplankton with their
prey, and can be used to set the zooplankton abundance relative to the 
phytoplankton.  We chose a value $A=0.02$ as bringing their abundance to within about
one order of magnitude.  Encounters scale with body mass with an exponent $\nu$;  
\citet{delong:12} recorded 95 \% confidence intervals $\approx 0.7 - 1.5$ 
for the exponent in a meta-analysis of protist-phytoplankton microcosms.
 
A predator-prey mass ratio was obtained from the relationship between ESD of 
predators and prey taken from \citet{wirtz:12}.  For unicellular plankton 
this is ESD$_P$ = 0.16 ESD$_Z$.  Scaling this from diameter to volume
and assuming neutral buoyancy gives a rounded value $\Delta = 5$.
\citet{wirtz:12} showed non-linearities at large zooplankton size (his Fig 4). Since 
our work was concerned with unicellular organisms, we worked with the left-hand 
end where the nonlinearity is small \citep[see eq. 8][]{wirtz:12}.   Here it is 
$\bar r$ that matters, estimated as $\bar r = 0.16$ \citep[][page 5]{wirtz:12}.
We varied the width of the feeding kernel to tune the connectance of 
the community. The value $\sigma = 1$ used here allows a high degree of 
connectedness, at the same time as ensuring prey cannot be large enough for 
cannibalism to take place. A food conversion efficiency $K = 0.6$ was assumed, 
following \citet{hartvig:11}.

The parameter $\chi = 0.4$ describing the form of density-dependent
predation was chosen by trial and error, as a small value that was often able to 
stabilise plankton assemblages.

All computations were carried with log transformed cell mass ($x=\textrm{ln}w$); 
this is because of large range of cell masses involved ($\sim$ 1 to 10$^6$).  
Computations were started by choosing species $w_i$s from uniformly-distributed 
random numbers over a range of approximately $\textrm{ln}2$ to 
$\textrm{ln}30000$ for phytoplankton, and $\textrm{ln}300$ to 
$\textrm{ln}(4\times10^6)$ for zooplankton. The random $w_i$s were rounded to
match the discretization of cell size used in the numerical integration. 
Eqs. \eqref{eq:MvFP}, \eqref{eq:MvFZ} were discretized into steps of width 
$\delta x$ as given in Table \ref{table:parameters}.  Integration of the full 
system was by the standard Euler method, with step sizes as small as computationally 
feasible. To obtain the equilibrium point, integration was carried out up 
to time $t_{max}$ to get close to equilibrium.  The equilibrium was found 
by the Newton-Raphson method; this uses the Jacobian matrix from which the 
leading eigenvalue at the equilibrium point was taken as the measure of local
asymptotic stability.


\bibliographystyle{apalike}
\bibliography{plankton5}

\begin{thebibliography}{}

\bibitem[Amato et~al., 2007]{amato:07}
Amato, A., Kooistra, W. H. C.~F., Ghiron, J. H.~L., Mann, D.~G., Pr{\"o}schold,
  T., and Montresor, M. (2007).
\newblock Reproductive isolation among sympatric cryptic species in marine
  diatoms.
\newblock {\em Protist}, 158:193--207.

\bibitem[Andersen and Beyer, 2006]{andersen:06}
Andersen, K.~H. and Beyer, J.~E. (2006).
\newblock Asymptotic size determines species abundance in the marine size
  spectrum.
\newblock {\em American Naturalist}, 168:54--61.

\bibitem[Armstrong, 1999]{armstrong:99}
Armstrong, R.~A. (1999).
\newblock Stable model structures for representing biogeochemical diversity and
  size spectra in plankton communities.
\newblock {\em Journal of Plankton Research}, 21:445--464.

\bibitem[Bell and Anderson, 1967]{bell:67}
Bell, G.~I. and Anderson, E.~C. (1967).
\newblock Cell growth and division {I}. {A} mathematical model with
  applications to cell volume distributions in mammalian suspension cultures.
\newblock {\em Biophysical Journal}, 7:329--351.

\bibitem[Beno{\^i}t and Rochet, 2004]{benoit:04}
Beno{\^i}t, E. and Rochet, M.-J. (2004).
\newblock A continuous model of biomass size spectra governed by predation and
  the effects of fishing on them.
\newblock {\em Journal of Theoretical Biology}, 226:9--21.

\bibitem[Blanchard et~al., 2005]{blanchard:05}
Blanchard, J.~L., Dulvy, N.~K., Jennings, S., Ellis, J.~R., Pinnegar, J.~K.,
  Tidd, A., and Kell, L.~T. (2005).
\newblock Do climate and fishing influence size-based indicators of {C}eltic
  {S}ea fish community structure?
\newblock {\em ICES Journal of Marine Science}, 62:405--411.

\bibitem[Boyce et~al., 2015]{boyce:15}
Boyce, D.~G., Frank, K.~T., and Leggett, W.~C. (2015).
\newblock From mice to elephants: overturning the ‘one size fits all’
  paradigm in marine plankton food chains.
\newblock {\em Ecology Letters}, 226:doi: 10.1111/ele.12434.

\bibitem[Capit{\'a}n and Delius, 2010]{capitan:10}
Capit{\'a}n, J.~A. and Delius, G.~W. (2010).
\newblock Scale-invariant model of marine population dynamics.
\newblock {\em Physical Review}, 81:061901.

\bibitem[Chen and Liu, 2010]{chen:10}
Chen, H. and Liu, H. (2010).
\newblock Relationships between phytoplankton growth and cell size in surface
  oceans: {I}nteractive effects of temperature, nutrients, and grazing.
\newblock {\em Limnology and Oceanography}, 55:965--972.

\bibitem[Chesson, 2000]{chesson:00}
Chesson, P. (2000).
\newblock Mechanisms of maintenance of species diversity.
\newblock {\em Annual Review of Ecology and Systematics}, 31:343--66.

\bibitem[Cuesta et~al., 2017]{cuesta:16}
Cuesta, J.~A., Delius, G.~W., and Law, R. (2017).
\newblock A size-disaggregated model for plankton dynamics.
\newblock {\em Journal of Mathematical Biology}, (to appear).

\bibitem[Datta et~al., 2010]{datta:10}
Datta, S., Delius, G.~W., and Law, R. (2010).
\newblock A jump-growth model for predator-prey dynamics: derivation and
  application to marine ecosystems.
\newblock {\em Bulletin of Mathematical Biology}, 72:1361--1382.

\bibitem[Datta et~al., 2011]{datta:11}
Datta, S., Delius, G.~W., Law, R., and Plank, M.~J. (2011).
\newblock A stability analysis of the power-law steady state of marine size
  spectra.
\newblock {\em Journal of Mathematical Biology}, 63:779--799.

\bibitem[DeLong et~al., 2010]{delong:10}
DeLong, J.~P., Okie, J.~G., Moses, M.~E., Sibly, R.~M., and Brown, J.~H.
  (2010).
\newblock Shifts in metabolic scaling, production, and efficiency across major
  evolutionary transitions of life.
\newblock {\em Proceedings of the National Academy of Sciences USA},
  107:12941--12945.

\bibitem[DeLong and Vasseur, 2012]{delong:12}
DeLong, J.~P. and Vasseur, D.~A. (2012).
\newblock Size-density scaling in protists and the links between
  consumer–resource interaction parameters.
\newblock {\em Journal of Animal Ecology}, 81:1193--120.

\bibitem[Diekmann et~al., 1983]{diekmann:83}
Diekmann, O., Lauwerier, H.~A., Aldenberg, T., and Metz, J. A.~J. (1983).
\newblock Growth, fission and the stable size distribution.
\newblock {\em Journal of Mathematical Biology}, 18:135--148.

\bibitem[Edwards and Bees, 2001]{edwards:01}
Edwards, A.~M. and Bees, M.~A. (2001).
\newblock Generic dynamics of a simple plankton population model with a
  non-integer exponent of closure.
\newblock {\em Chaos, Solitons and Fractals}, 12:289--300.

\bibitem[Flynn et~al., 2013]{flynn:13}
Flynn, K.~J., Stoeker, D.~K., Mitra, A., Raven, J.~A., Glibert, P.~M., Hansen,
  P.~J., Gran{\'e}li, E., and Burkholder, J.~M. (2013).
\newblock Misuse of the phytoplankton – zooplankton dichotomy: the need to
  assign organisms as mixotrophs within plankton functional types.
\newblock {\em Journal of Plankton Research}, 35:3--11.

\bibitem[Fox, 2013]{fox:13}
Fox, J.~W. (2013).
\newblock The intermediate disturbance hypothesis should be abandoned.
\newblock {\em Trends in Ecology \& Evolution}, 28:86--92.

\bibitem[Fredrickson et~al., 1967]{fredrickson:67}
Fredrickson, A.~G., Ramakrishna, D., and Tsuchiya, H.~M. (1967).
\newblock Statistics and dynamics of procaryotic cell populations.
\newblock {\em Mathematical Biosciences}, 1:327--374.

\bibitem[Fulton, 2010]{fulton:10}
Fulton, E.~A. (2010).
\newblock Approaches to end-to-end ecosystem models.
\newblock {\em Journal of Marine Systems}, 81:171--183.

\bibitem[Gaedke, 1992]{gaedke:92}
Gaedke, U. (1992).
\newblock The size distribution of plankton biomass in a large lake and its
  seasonal variability.
\newblock {\em Limnology and Oceanography}, 37:1202--1220.

\bibitem[Giometto et~al., 2013]{giometto:13}
Giometto, A., Altermatt, F., Carrara, F., Maritan, A., and Rinaldo, A. (2013).
\newblock Scaling body size fluctuations.
\newblock {\em Proceedings of the National Academy of Sciences USA},
  110:4646--4650.

\bibitem[Guiet et~al., 2016]{guiet:16}
Guiet, J., Poggiale, J.-C., and Maury, O. (2016).
\newblock Modelling the community size-spectrum: recent developments and new
  directions.
\newblock {\em Ecological Modelling}, 337:4--14.

\bibitem[Hansen et~al., 1994]{hansen:94}
Hansen, B., Bjornsen, P.~K., and Hansen, P.~J. (1994).
\newblock The size ratio between planktonic predators and their prey.
\newblock {\em Limnology and Oceanography}, 39:395--403.

\bibitem[Hartvig et~al., 2011]{hartvig:11}
Hartvig, M., Andersen, K.~H., and Beyer, J.~E. (2011).
\newblock Food web framework for size-structured populations.
\newblock {\em Journal of Theoretical Biology}, 272:113--122.

\bibitem[Heath, 1995]{heath:95}
Heath, M.~R. (1995).
\newblock Size spectrum dynamics and the planktonic ecosystem of {L}och
  {L}innhe.
\newblock {\em ICES Journal of Marine Science}, 52:627--642.

\bibitem[Heijmans, 1984]{Heijmans:84}
Heijmans, H. J. A.~M. (1984).
\newblock On the stable size distribution of populations reproducing by fission
  into two unequal parts.
\newblock {\em Mathematical Biosciences}, 72:19--50.

\bibitem[Henson, 2003]{henson:03}
Henson, M.~A. (2003).
\newblock Dynamic modeling of microbial cell populations.
\newblock {\em Current Opinion in Biotechnology}, 14:460--467.

\bibitem[Holling, 1959]{holling:59}
Holling, C.~S. (1959).
\newblock Some characteristics of simple types of predation and parasitism.
\newblock {\em Canadian Entomologist}, 91:385--398.

\bibitem[Huisman and Weissing, 1999]{huisman:99}
Huisman, J. and Weissing, F.~J. (1999).
\newblock Biodiversity of plankton by species oscillations and chaos.
\newblock {\em Nature}, 402:407--410.

\bibitem[Hutchinson, 1961]{hutchinson:61}
Hutchinson, G.~E. (1961).
\newblock The paradox of the plankton.
\newblock {\em American Naturalist}, 95:137--145.

\bibitem[Jezbera et~al., 2006]{jezbera:06}
Jezbera, J., Hor\v{n}\'{a}k, K., and \v{S}imek, K. (2006).
\newblock Prey selectivity of bacterivorous protists in different size
  fractions of reservoir water amended with nutrients.
\newblock {\em Environmental Microbiology}, 8:1330--1339.

\bibitem[Kacelnik et~al., 1992]{kacelnic:92}
Kacelnik, A., Krebs, J.~R., and Bernstein, C. (1992).
\newblock The ideal free distribution and predator-prey populations.
\newblock {\em Trends in Ecology and Evolution}, 7:50--55.

\bibitem[Kempes et~al., 2012]{kempes:12}
Kempes, C.~P., Dutkiewicz, S., and Follows, M.~J. (2012).
\newblock Growth, metabolic partitioning, and the size of microorganisms.
\newblock {\em Proceedings of the National Academy of Sciences USA},
  109:495--500.

\bibitem[Leibold, 1996]{leibold:96}
Leibold, M.~A. (1996).
\newblock Graphical model of keystone predators in food webs: trophic
  regulation of abundance, incidence, and diversity patterns in communities.
\newblock {\em American Naturalist}, 147:784--812.

\bibitem[L{\'o}pez-Sandoval et~al., 2014]{lopez:14}
L{\'o}pez-Sandoval, D.~C., Rodr{\'i}guez-Ramos, T., Cerme{\~n}o, P., Sobrino,
  C., and Mara{\~n\'o}n, E. (2014).
\newblock Photosynthesis and respiration in marine phytoplankton:
  {R}elationship with cell size, taxonomic affiliation, and growth phase.
\newblock {\em Journal of Experimental Marine Biology and Ecology},
  457:151--159.

\bibitem[Mara{\~n\'o}n et~al., 2013]{maranon:13}
Mara{\~n\'o}n, E., Cerme{\~n}o, P., L{\'o}pez-Sandoval, D.~C.,
  Rodr{\'{\i}}guez-Ramos, T., Sobrino, C., Huete-Ortega, M., Blanco, J.~M., and
  Rodr{\'{\i}}guez, J. (2013).
\newblock Unimodal size scaling of phytoplankton growth and the size dependence
  of nutrient uptake and use.
\newblock {\em Ecology Letters}, 16:371--379.

\bibitem[May, 1972]{may:72}
May, R.~M. (1972).
\newblock Will a large complex system be stable?
\newblock {\em Nature}, 238:413--414.

\bibitem[Mc{M}anus and Katz, 2009]{mcmanus:09}
Mc{M}anus, G.~B. and Katz, L.~A. (2009).
\newblock Molecular and morphological methods for identifying plankton: what
  makes a successful marriage?
\newblock {\em Journal of Plankton Research}, 31:1119--1129.

\bibitem[Morozov, 2010]{morosov:10}
Morozov, A.~Y. (2010).
\newblock Emergence of {H}olling type {III} zooplankton functional response:
  bringing together field evidence and mathematical modelling.
\newblock {\em Journal of Theoretical Biology}, 265:45--54.

\bibitem[Nachman, 2006]{nachman:06}
Nachman, G. (2006).
\newblock A functional response model of a predator population foraging in a
  patchy habitat.
\newblock {\em Journal of Animal Ecology}, 75:948--958.

\bibitem[Plank and Law, 2011]{plank:11}
Plank, M.~J. and Law, R. (2011).
\newblock Ecological drivers of stability and instability in marine ecosystems.
\newblock {\em Theoretical Ecology}, 5:465--480.

\bibitem[Quinones et~al., 2003]{quinones:03}
Quinones, A., Platt, T., and Rodr{\'i}guez, J. (2003).
\newblock Patterns of biomass-size spectra from oligotrophic waters of the
  {N}orthwest {A}tlantic.
\newblock {\em Progress in Oceanography}, 57:405--427.

\bibitem[Roy and Chattopadhyay, 2007]{roy:07}
Roy, S. and Chattopadhyay, J. (2007).
\newblock Towards a resolution of ‘the paradox of the plankton’: a brief
  overview of the proposed mechanisms.
\newblock {\em Ecological Complexity}, 4:26--33.

\bibitem[San~Martin et~al., 2006]{sanmartin:06}
San~Martin, E., Irigoien, X., Harris, R.~P., L{\'o}pez-Urrutia, {\'A}., Zubkov,
  M.~Z., and Heywood, J.~L. (2006).
\newblock Variation in the transfer of energy in marine plankton along a
  productivity gradient in the {A}tlantic {O}cean.
\newblock {\em Limnology and Oceanography}, 51:2084--2091.

\bibitem[Scheffer et~al., 2003]{scheffer:03}
Scheffer, M., Rinaldi, S., Huisman, J., and Weissing, F.~J. (2003).
\newblock Why plankton communities have no equilibrium: solutions to the
  paradox.
\newblock {\em Hydrobiologia}, 491:9--18.

\bibitem[Sheldon et~al., 1972]{sheldon:72}
Sheldon, R., Prakash, A., and Sutcliffe~Jr., W.~H. (1972).
\newblock The size distribution of particles in the ocean.
\newblock {\em Limnology and Oceanography}, 17:327--340.

\bibitem[Silvert and Platt, 1978]{silvert:78}
Silvert, W. and Platt, T. (1978).
\newblock Energy flux in the pelagic ecosystem: a time-dependent equation.
\newblock {\em Limnology and Oceanography}, 23:813--816.

\bibitem[Silvert and Platt, 1980]{silvert:80}
Silvert, W. and Platt, T. (1980).
\newblock In Kerfoot, W.~C., editor, {\em Evolution and ecology of zooplankton
  communities}, chapter Dynamic energy-flow model of the particle size
  distribution in pelagic ecosystems, pages 754--763. University Press of New
  England, Hanover, New Hampshire.

\bibitem[Sinko and Streifer, 1971]{sinko:71}
Sinko, J.~W. and Streifer, W. (1971).
\newblock A model for populations reproducing by fission.
\newblock {\em Ecology}, 52:330--335.

\bibitem[Sprules et~al., 1991]{sprules:91}
Sprules, W.~G., Brandt, S.~B., Stewart, D.~J., Munawar, M., Jin, E.~H., and
  Love, J. (1991).
\newblock Biomass size spectrum of the {L}ake {M}ichigan pelagic food web.
\newblock {\em Canadian Journal of Fisheries and Aquatic Sciences},
  48:105--115.

\bibitem[Tang, 1995]{tang:95}
Tang, E. P.~Y. (1995).
\newblock The allometry of algal growth rates.
\newblock {\em Journal of Plankton Research}, 17:1325--1335.

\bibitem[Thingstad and Lignell, 1997]{thingstad:97}
Thingstad, T.~F. and Lignell, R. (1997).
\newblock Theoretical models for the control of bacterial growth rate,
  abundance, diversity and carbon demand.
\newblock {\em Aquatic Microbial Ecology}, 13:19--27.

\bibitem[Truscott and Brindley, 1994]{truscott:94}
Truscott, J.~E. and Brindley, J. (1994).
\newblock Ocean plankton populations as excitable media.
\newblock {\em Bulletin of Mathematical Biology}, 56:981--998.

\bibitem[Va\r{g}e et~al., 2014]{vage:14}
Va\r{g}e, S., Storesund, J.~E., Giske, J., and Thingstad, T.~F. (2014).
\newblock Optimal defense strategies in an idealized microbial food web under
  trade-off between competition and defense.
\newblock {\em PLOS ONE}, 9:e101415.

\bibitem[von Bertalanffy, 1957]{vonbertalanffy:57}
von Bertalanffy, L. (1957).
\newblock Quantitative laws in metabolism and growth.
\newblock {\em Quarterly Review of Biology}, 32:217--231.

\bibitem[Winter et~al., 2010]{winter:10}
Winter, C., Bouvier, T., Weinbaue, M.~G., and Thingstad, T.~F. (2010).
\newblock Trade-offs between competition and defense specialists among
  unicellular planktonic organisms: the ``killing the winner'' hypothesis
  revisited.
\newblock {\em Microbial and Molecular Biology Reviews}, 74:42--57.

\bibitem[Wirtz, 2012]{wirtz:12}
Wirtz, K.~W. (2012).
\newblock Who is eating whom? {M}orphology and feeding type determine the size
  relation between planktonic predators and their ideal prey.
\newblock {\em Marine Ecology Progress Series}, 445:1--12.

\bibitem[Yurista et~al., 2014]{yurista:14}
Yurista, P.~M., Yule, D.~L., Balge, M., Van{A}lstine, J.~D., Thompson, J.~A.,
  Gamble, A.~E., Hrabik, T.~R., Kelly, J.~R., Stockwell, J.~D., and Vinson,
  M.~R. (2014).
\newblock A new look at the {L}ake {S}uperior biomass size spectrum.
\newblock {\em Canadian Journal of Fisheries and Aquatic Sciences},
  71:1324--133.

\end{thebibliography}

\end{document}